\documentstyle[11pt,psfig,oldlfont]{article}
\def\cm{\,{\rm cm}}
\def\K{\,{\rm K}}

\def\erg{\,{\rm erg}}

\def\km{\,{\rm km}}

\def\kpc{\,{\rm kpc}}
\def\pc{\,{\rm pc}}

\def\sr{\,{\rm sr}}
\def\s{\,{\rm s}}

\def\Hz{\,{\rm Hz}}

\def\aua#1#2{{ #1, }{A\&A,}{ #2}}

\def\apj#1#2{{#1, }{ApJ,} { #2}}
\def\apjs#1#2{{#1, }{ApJS,} { #2}}
\def\aj#1#2{{#1, }{AJ,} { #2}}
\def\mnras#1#2{{#1, }{MNRAS,} { #2}}

\def\Atoday{\ifcase\month\or
  January\or February\or March\or April\or May\or June\or
  July\or August\or September\or October\or November\or December\fi
  \space\number\day, \number\year}
\def\Etoday{\number\day\space\ifcase\month\or
  January\or February\or March\or April\or May\or June\or
  July\or August\or September\or October\or November\or December\fi
  \space\number\year}

\immediate\write16{... Special symbols are defined}
\def\la{\mathrel{\hbox{\rlap{\hbox{\lower4pt\hbox{$\sim$}}}\hbox{$<$}}}}
\def\ga{\mathrel{\hbox{\rlap{\hbox{\lower4pt\hbox{$\sim$}}}\hbox{$>$}}}}
\def\lse{\mathrel{\hbox{\rlap{\hbox{\raise4pt\hbox{$\<$}}}\hbox{$\simeq$}}}}
\def\gse{\mathrel{\hbox{\rlap{\hbox{\raise4pt\hbox{$\>$}}}\hbox{$\simeq$}}}}
\def\loa{\mathrel{\hbox{\rlap{\hbox{\lower4pt\hbox{$\approx$}}}\hbox{$<$}}}}
\def\goa{\mathrel{\hbox{\rlap{\hbox{\lower4pt\hbox{$\approx$}}}\hbox{$>$}}}}

\def\ed{\end{document}}

\def\beq#1{\begin{equation}\label{#1}}
\def\eeq{\end{equation}}
\def\beqa#1{\begin{eqnarray}\label{#1}}
\def\eeqa{\end{eqnarray}}

\def\bfig{\begin{figure}[h] \centerline{\hbox{}}\vfill}
\def\efig{\end{figure}\vfill\newpage}

\def\spose#1{\hbox to 0pt{#1\hss}}
\def\simlt{\mathrel{\spose{\lower 3pt\hbox{$\mathchar"218$}}
     \raise 2.0pt\hbox{$\mathchar"13C$}}}
\def\simgt{\mathrel{\spose{\lower 3pt\hbox{$\mathchar"218$}}
     \raise 2.0pt\hbox{$\mathchar"13E$}}}
\def\simpropto{\mathrel{\spose{\lower 3pt\hbox{$\mathchar"218$}}
     \raise 2.0pt\hbox{$\propto$}}}

\def\addr{\footnotesize\it}

\newcommand{\op}{Ly$\alpha$\ }

\begin{document}

\begin{titlepage}   


\begin{center}
\bigskip

{\bf \Large 
Non-equilibrium effects on line-of-sight 
size estimates of QSO absorption systems}

\bigskip

{\large Martin G. Haehnelt$^{1}$, Michael Rauch$^{2,3}$ \& Matthias
 Steinmetz$^{1,4}$  }
\bigskip

{\addr $^{1}$ Max-Planck-Institut f\"ur Astrophysik,
Karl-Schwarzschild-Stra\ss e 1,\\
D-85740 Garching b. M\"unchen, Germany\hfill}\\
{\addr $^{2}$Astronomy Dept., 105-24 California Institute of
Technology, Pasadena 91125, USA\hfill}\\
{\addr $^{3}$Hubble fellow\hfill}\\
{\addr $^{4}$present address: Dept. of Astronomy, UC Berkeley
94720, USA\hfill}\\
\vskip 0.5 cm

{\addr e-mail: mhaehnelt@mpa-garching.mpg.de, mr@astro.caltech.edu,
mhs@astro.berkeley.edu}

\vskip 1.5cm
{Subject Headings:  cosmology: theory, observations --- galaxies:
formation, evolution --- intergalactic medium --- quasars: absorption lines} 
\vskip 1.5cm
\vskip 1.5cm

\end{center}

\abstract{Estimates of the linear extent of
heavy-element absorption systems along the line-of-sight 
to a QSO often assume that the cloud is photoionized and that the 
temperature takes the equilibrium value where photo-heating 
balances line cooling. We show that rather small deviations from this
photoionization equilibrium temperature caused by additional heating 
processes will lead to an overestimate of the neutral hydrogen
fraction and thus to an underestimate of the thickness of the absorber by
about two orders of magnitude. Such temperature deviations are  
indicated both by observations and numerical simulations.  
This interpretation reconciles the  discrepancy between  
the rather small extent of heavy-element-absorption systems
parallel to the line-of-sight obtained from a standard photoionization 
analysis  and  the much larger transverse sizes estimates
inferred from the observation of common absorption in the spectra of
close quasar pairs.}
\end{titlepage}

\section{Introduction}

Recently new size estimates for QSO absorption systems
have modified our view of the physical properties of low and 
intermediate column density QSO absorption
systems. Dinshaw et al. (1994) and Bechthold et al. (1994) have observed
coincident absorption lines in the spectra of the
close  quasar pair 1343+26 at redshift
$z=2.03$  with angular separation 9.5 arcmin (corresponding to a proper 
separation of $40h^{-1} kpc$ for $q_{0} = 0.5$). Assuming a population
of spherical absorbers of fixed size the authors
of both studies inferred a typical absorber radius of 
$\sim 100 h^{-1}\kpc$.
Further observations of a few other quasar pairs  confirmed the
new size estimates which are  at least an order of magnitude 
larger than assumed previously. Even so the assumption of  
spherical absorbers of fixed size is rather  simplistic,   
more sophisticated modeling gives similar results with 
inferred sizes even larger by a factor $1.5-2$ (Charlton et al. 1996).

If the typical values measured for the transverse 
size were also characteristic for the 
extent parallel to the line-of-sight (LOS) then the usually assumed fiducial
values for  the total  density and  the neutral hydrogen fraction 
would have to be revised by factors $3-30$. 
However, Rauch \& Haehnelt (1995) have argued 
that the absorbers are flattened structures in which case more
moderate correction factors  apply.

To further probe the nature of the absorption systems more detailed 
information on the  physical properties and especially 
the absorber extent parallel to  the LOS are necessary. 
Low-and intermediate column density absorption systems are 
known to be highly ionized and photoionization is generally 
assumed to be the ionizing mechanism (Bergeron \& Stasi\'nska 1986).  
If the absorbers were in full thermal photoionization equilibrium  and
strength and spectrum of the ionizing background radiation were
known, a determination of the ionization  parameter would be
sufficient to determine the mean total density, the
neutral fraction of the absorbing gas and thus its extent
parallel to the  the LOS (e.g. Giallongo \& Petitjean 1994).
The recent work by Cowie et al. (1995)  used both the 
CII/CIV column density ratios of the associated metal absorption 
and the  temperature of the clouds inferred  from their Doppler
parameter to determine  the ionization 
parameter. Using standard photoionization 
models Cowie et al. inferred a thickness of $100 \pc$
for the CIV absorbing gas clouds associated with low column density 
absorption systems. Considering the  measured transverse sizes
obtained for only slightly larger column density systems 
this result seems quite puzzling.  It would imply  axis ratios of 
$10^{-3} - 10^{-4}$ for a homogeneous distribution of CIV absorbing
gas. Such a configuration should be quite unstable. Problems of an
alternative  model where the  CIV absorbing gas is highly
inhomogeneous and consists  of many small clumps with 
covering factor of order unity have been discussed  by Mo (1994) 
and Mo \& Miralda-Escud\`e (1996).  We suggest here a 
different solution  to reconcile the apparent large difference 
between the extent parallel and transverse to the LOS. 
In  section 2 and 3 we will show that full thermal photoionization  
equilibrium is unlikely to be a valid assumption and   
will investigate non-equilibrium effects on the ionization state 
of the gas. In section 4 we  discuss implications for the inferred 
thickness of the absorbers. Section 5 contains the conclusions.

\section{Non-equilibrium effects on the ionization state of 
carbon  and hydrogen}

Observations and numerical simulation indicate that the temperature of
the absorbing gas does not take the equilibrium  value where
photo-heating balances line cooling (a situation which we will
call {\it full} (=thermal photoionization) equilibrium in the following). 
This is already seen in the
observed b-value distribution of \op lines. There is  a broad  range
of b-values between $20$ and $100\km \s^{-1}$ (Hu et al. 1995).  
This corresponds to a temperature range between $1.5 \times 10^{4} \K$
and $3 \times 10^{5} \K$  if the line profiles are due to thermal
broadening. High resolution studies of metal absorption
lines do indeed indicate that a single temperature may not be a good
approximation.
Rauch et al. (1996) used a high resolution sample of 
CIV and SIV lines to show that the turbulent contribution 
to the Doppler broadening is small (considerably less than $10 \km \s^{-1}$)
and that there is a range of gas temperatures  between $2\times 10^{4}
\K$ and $3\times 10^{5}\K$ with a mean temperature of $4\times 10^{4} \K$. 
As discussed by Haehnelt, Steinmetz \& Rauch (1996) and shown in
Fig. 1 numerical simulations exhibit  a similar range of gas temperatures.
The deviations from full  equilibrium are   easy to explain  as the 
equilibrium models neglect shock heating due to the collapse of
structures. In the following we will discuss the effect of such 
deviations on the ionization state of the gas.   The simulations  also
show a  small turbulent contribution to the line width at the relevant
column densities.

As pointed out by Cowie et al. (1995)
the CIV $\lambda\lambda 1548,1550$ and CII $\lambda 1334$ seems the
most promising line pair for a reliable determination of the ionization
parameter of low and intermediate column density absorption systems. 
In the following we will therefore concentrate on hydrogen
and carbon.  As we are only interested in absorbers of moderate column
density we will neglect self-shielding and radiative transfer effects.
We have used the photoionization code CLOUDY 
(Ferland 1993) to calculate the ionization state  for a grid of  
ionization parameter and temperatures dropping the assumption of 
full photoionization equilibrium. CLOUDY was used in 
its ``\op absorber mode''.  The absorber is then assumed to be an
infinite homogeneous slab of gas of low metallicity, optically thin to 
ionizing radiation and illuminated from both sides 
by a homogeneous UV field. $I_{\nu} \propto \nu^{-\alpha}$
was adopted for the spectrum of the ionizing photons.
The ionization parameter  
\begin{equation} 
U = 4.2\times 10^{-5}\; (I_{21}/n_{H}), 
\end{equation}
where 
$I_{21}$ is the ionizing flux in units of
 $10^{-21} \erg \cm^{-2} \s^{-1}  \Hz^{-1} \sr^{-1}$,
was defined as  in Donahue \& Shull (1991) with $\alpha = 1.5$. 

In figure 2 the fraction of CIV, CII and HI is shown  as function of
ionization parameter. The solid curves assume full thermal
photoionization equilibrium. 
While for large ionization parameter the  temperature dependence of 
the CII and CIV fraction is weak,  it is rather 
strong for small ionization parameter ($U\la 0.01$). The
temperature dependence of the HI fraction is generally strong  
with an increase towards low ionization parameter/high densities. The 
ion abundances derived from the equilibrium and non-equilibrium cases
can differ by several orders of magnitude.

\section{Ionization parameter and total density}

Figure 3 shows how the ionization parameter inferred 
from the CII/CIV ratio changes if we drop the assumption 
of full  equilibrium. The dependence on 
temperature is weak as long as the temperature does not exceed 
a  critical value, beyond which CII/CIV becomes virtually independent of 
the ionizing parameter. For  observed  CII/CIV ratios this critical 
temperature  is about $10^{5} K$, somewhat above the temperature 
indicated by  the Doppler parameter for most of the carbon lines
(Rauch et al. 1996). The observed CII/CIV ratios  will therefore give 
ionization parameter which are correct to a factor of about three 
if full equilibrium is assumed, although  there are some  
additional uncertainties related to the unknown spectrum of the
ionizing radiation.  Hu et al. (1995) used the strong dependence of the
ionization parameter on temperature in full equilibrium models as a 
second independent method to determine the ionization parameter. 
However, the  typical  temperature distribution of  gas particles 
in a SPH simulation shown in Figure 1 (Steinmetz 1996, 
Haehnelt et al. 1996)
indicates that a strong correlation between temperature and 
ionization parameter should not be expected. The temperature 
of full equilibrium models is therefore probably not a good indicator 
of the ionization parameter.

\section{Neutral hydrogen fraction and size estimation}

Figure 2a shows that the neutral hydrogen fraction is strongly
temperature dependent for fixed ionization parameter.
Even if we determine the ionization parameter to about a factor of three 
a determination of the temperature is absolutely essential
to fix the neutral hydrogen fraction $x = n_{HI} /n_{H}$.
The extent of the absorber parallel to the LOS is given by 

\begin{equation} 
D =  \frac{N_{HI}}{x\,n_{H}}. 
\end{equation}

Any error in the determination of the neutral fraction 
therefore directly propagates into an error in the determination
of $D$. In the upper panel of Figure 4 the inferred 
extent parallel to the LOS 
is shown as a function of the CII/CIV ratio for 
$\alpha = 1.5$.  The dashed curves are for different temperatures
as indicated  on the plot. The left axis is for a column density 
of $10^{14} \cm^{-2}$ and the right axis is for a column density of 
$10^{16} \cm^{-2}$. The dashed line assumes full  equilibrium.
The inferred size is extremely sensitive to the temperature. An increase
in the assumed temperature by only a factor of 2.5 from
$2 \times 10^{4}\K$ to $5 \times 10^{4}\K$ increases the inferred 
extent along the LOS by a factor of about 30.

\section{Discussion and Conclusions} 

Cowie et al.(1995) used high resolution spectra obtained with the
HIRES spectrograph of the Keck telescope to determine the 
CII/CIV column  density ratio  of low and intermediate 
column density absorbers. They derived an upper limit of 
$N_{CII}/N_{CIV} \le 0.1$ for  absorption systems with neutral 
hydrogen column  densities of $10^{14} \cm^{-2}$
and found $N_{CII}/N_{CIV} \sim 0.1-0.3$ for larger column
densities. From this they inferred  $U\ga 10^{-2}$  
and $U\sim 10^{-3} - 10^{-2.5}$, respectively, 
using the full equilibrium photoionization models of Donahue \& Shull
(1991). As discussed in section 2.3 these estimates do not change much if the 
assumption of full thermal photoionization equilibrium is dropped. 
However, numerical  simulation indicate that typical ionization 
parameter of low column density systems are higher by at least an order of 
magnitude than the observed lower limits. This corresponds to  
$N_{CII}/N_{CIV} \sim 10^{-3}-10^{-2}$ and it seems unlikely that
CII will be actually detected in low-column density absorption systems.

Following Cowie et al. 1995 and assuming a full equilibrium model 
with $N_{CII}/N_{CIV} \sim 0.1$ we would infer an extent along the 
LOS of $\sim 100 \pc$  for an absorber with  $N_{H}=10^{14} \cm^{-2}$ 
in good agreement with their result. Dropping the assumption that 
the temperature takes the equilibrium value and taking into account 
that we have only a upper limit for $N_{CII}/N_{CIV}$, the inferred 
extent is increased to at least $10-100 \kpc$. For larger column 
densities ($N_{HI} \sim 10^{16} \cm^{-2}$) with actually observed 
$N_{CII}/N_{CIV} \sim 0.1-0.3$
the extent along the LOS changes less dramatically, 
from 1 to 10 kpc. The sizes inferred from our  non-equilibrium 
calculations are in good agreement with the sizes of CIV absorbing
regions which we found recently in numerical  galaxy formation 
simulations (Haehnelt, Steinmetz \& Rauch 1996).

If the ionizing spectrum were softer the  
size estimates would be even larger. This is shown in Figure 4b which 
assumes a spectral index $\alpha =5$. For this very soft spectrum the 
inferred extent becomes unreasonable large. It reflects the 
well known fact that a stellar-like spectrum is not able to reproduce  
observed line ratios (e.g. Steidel 1990). Taking into account the presence 
of a  He absorption edge at 4 Rydberg in the spectrum of the UV background
has a more moderate effect on the size estimates, increasing them
by a factor of about two for a depression of the spectrum 
by two orders of magnitude at 4 Rydberg.

Dropping the assumption of a homogeneous single-temperature 
absorbing medium would further increase the size estimates. 
If there were either a temperature and density gradient towards the 
centre of the absorber (as seen  in the numerical simulations) or if the
medium were clumpy,  CII absorption would be biased towards  
the cooler and denser regions. The assumption of a homogeneous
absorber then overestimates the HI absorption-weighted 
CII/CIV ratio of the gas.

We conclude that equilibrium photoionization models considerably
underestimate the size of the absorption systems parallel to the LOS
if the temperature of the absorbing gas is higher than the equilibrium
value where photo-heating balances line cooling. 
Such departures from equilibrium temperatures are indeed indicated both by
observations and numerical simulations. The large discrepancy between the
size estimates parallel and transverse to the LOS
then disappears, reducing the need for models in which a population of
numerous tiny clouds (R$\sim$ 100 pc) is kept in place by much larger
halos.


\noindent 
\section {Acknowledgments}
We thank 
Gary Ferland for making CLOUDY available to us. 
MR is grateful to NASA for support through
grant HF-01075.01-94A from the Space Telescope Science Institute, 
which is operated by the Association of Universities for Research in
Astronomy, Inc., under NASA contract NAS5-26555.
Support by NATO grant CRG 950752 and the ``Sonderforschungsbereich 375-95 
f\"ur Astro-Teilchenphysik der Deutschen  Forschungsgemeinschaft'' 
is  gratefully acknowledged. 
We also thank the referee Megan Donahue for a  helpful report.

\def\rf{\par \noindent \hangindent=3pc \hangafter=1}

\setcounter{secnumdepth}{-1}
\section{References}

\rf Bechtold J., Crotts A.P.S., Duncan R.S., Fang Y., \apj{1994}{437}, L83


\rf Bergeron J., Stasi\'nska G., \aua{1986}{169}, 1



\rf Charlton J.C., Anninos P., Yu Zh., Norman M., 1996, submitted to ApJ, 
astro-ph/9601152

\rf Cowie L.L., Songaila A., Kim T.-S., Hu E. \aj{1995}{109}, 1522

\rf Dinshaw N., Impey, C.D., Foltz C.B., Weymann R.J., Chaffee  F.H.,
\apj{1994}{437}, L87

\rf Donahue M., Shull J.M., \apj{1991}{383}, 511


\rf Ferland G.J., 1993, University of Kentucky Department of Physics and 
Astronomy Internal Report

\rf Giallongo E., Petitjean P., \apj{1994}{426}, L61

\rf Haehnelt M., Steinmetz M., Rauch M.,  \apj{1996}{465}, L95 

\rf Hu E.M., Kim T.-S., Cowie L., Songaila A., Rauch, M., \aj{1995}{110}, 1526
 









\rf Mo H.J., \mnras{1994}{269}, L49

\rf Mo H.J., Miralda-Escud\'e J., 1996, ApJ, in press, astro-ph/9603027

 


\rf Rauch M., Haehnelt M., \mnras{1995}{275}, L76  

\rf Rauch M., Sargent W.L.W., Womble D.S., Barlow, T.A., 1996, ApJ,
    in press, astro-ph/960604


\rf Steidel C.C., \apjs{1990}{74}, 37 


\rf Steinmetz M., \mnras{1996}{278}, 1005






\vfill
\break

\clearpage

\newpage

\begin{figure}[p]
\null
\vspace{-2.0cm}
\psfig{file=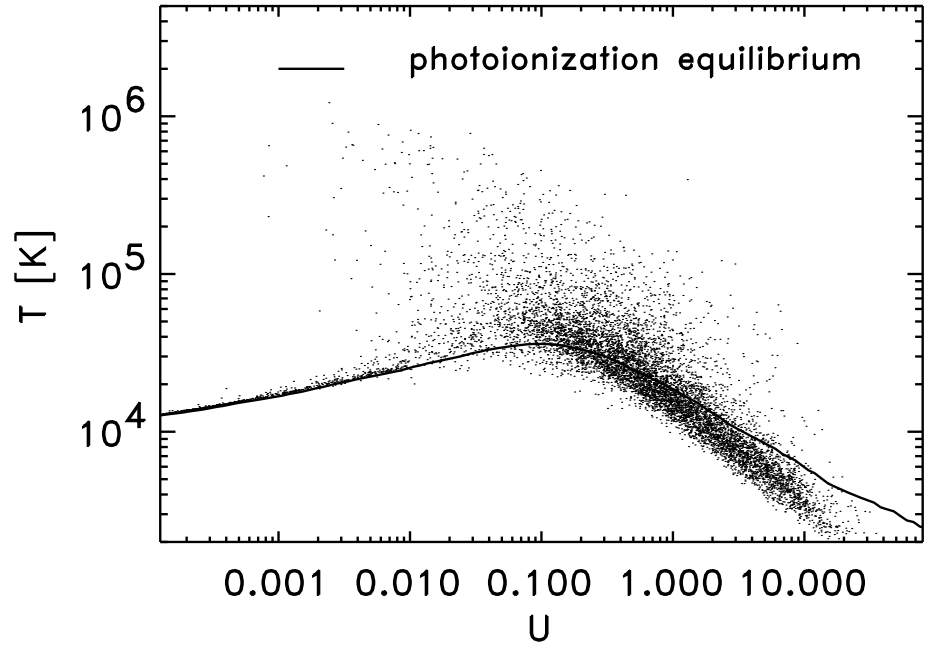,width=14.cm}
\vspace{2.0cm}
\caption{
 A typical  temperature/ionization parameter  distribution of gas
particles in a SPH simulation at redshift $z=3$ is shown. 
The ionizing background ($I_{\nu} \propto \nu^{-\alpha}$, $\alpha = 1.5$, 
$I_{21} (z=3) = 0.3$) is fixed, while the density varies. 
The effects of Compton cooling due to the cosmic microwave background, 
adiabatic cooling and shock heating are included (see also 
Haehnelt, Steinmetz \& Rauch 1996). 
The solid curve shows the corresponding relation assuming full 
thermal photoionization equilibrium.}
\end{figure}

\newpage

\begin{figure}[p]
\null
\vspace{-2.0cm}
\centerline{
\psfig{file=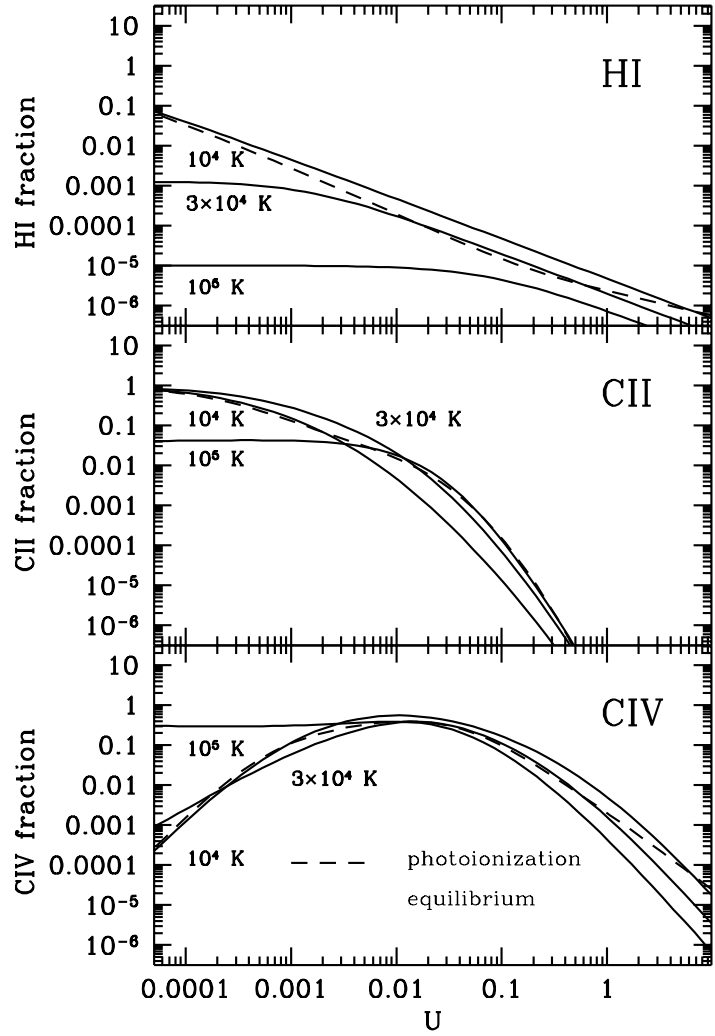,width=16.cm}}
\vspace{-0.0cm}
\caption{
The three panels show the  fraction of hydrogen and carbon in the form 
of HI, CII, and CIV  as a function  of the ionization parameter U 
for three different  temperatures as indicated on the plot
($I_{\nu} \propto \nu^{-\alpha}$, $\alpha = 1.5$). The dashed curve
assumes that the temperature takes the equilibrium value where photo-heating 
balances line cooling.}
\end{figure}

\newpage

\begin{figure}[p]
\vspace{2.0cm}
\centerline{
\psfig{file=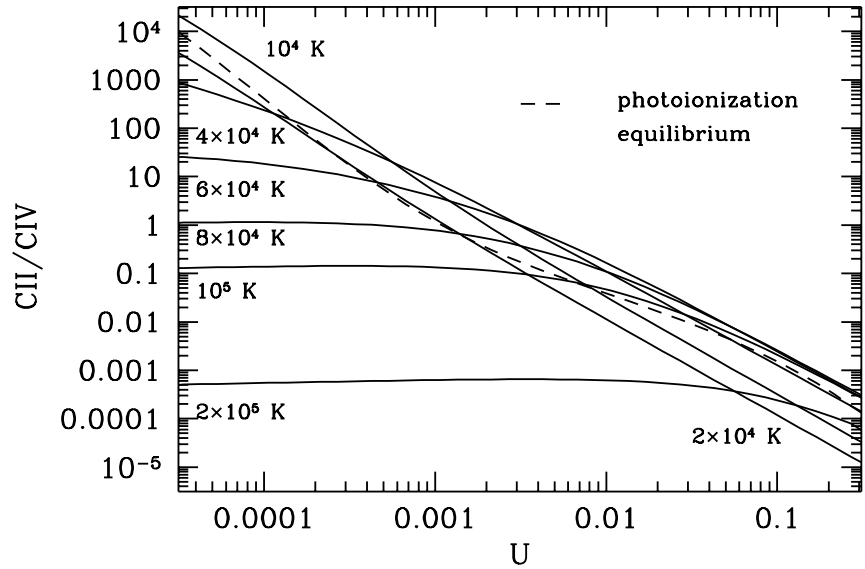,width=14.cm}}
\vspace{-4.0cm}
\caption{
The solid curves  show the CII/CIV ratio as  a function  
of the ionization parameter U  for different temperatures as indicated
on the plot ($I_{\nu} \propto \nu^{-\alpha}$, $\alpha = 1.5$). 
The dashed curve assumes that the temperature takes the equilibrium
value where photo-heating balances line cooling}.
\end{figure}

\newpage

\begin{figure}[p]
\null

\vspace{-2.0cm}
\psfig{file=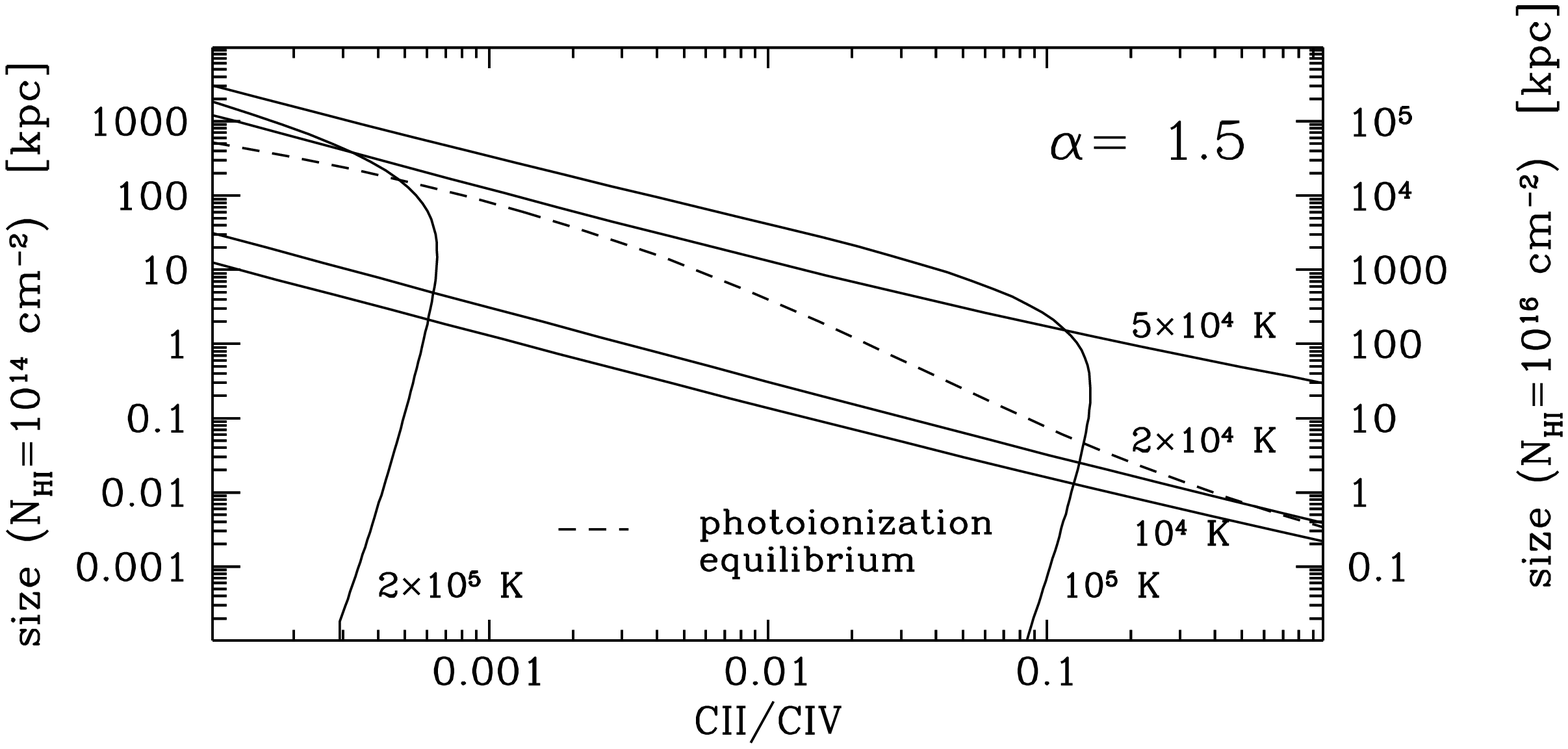,width=14.cm}

\vspace{-2.0cm}
\psfig{file=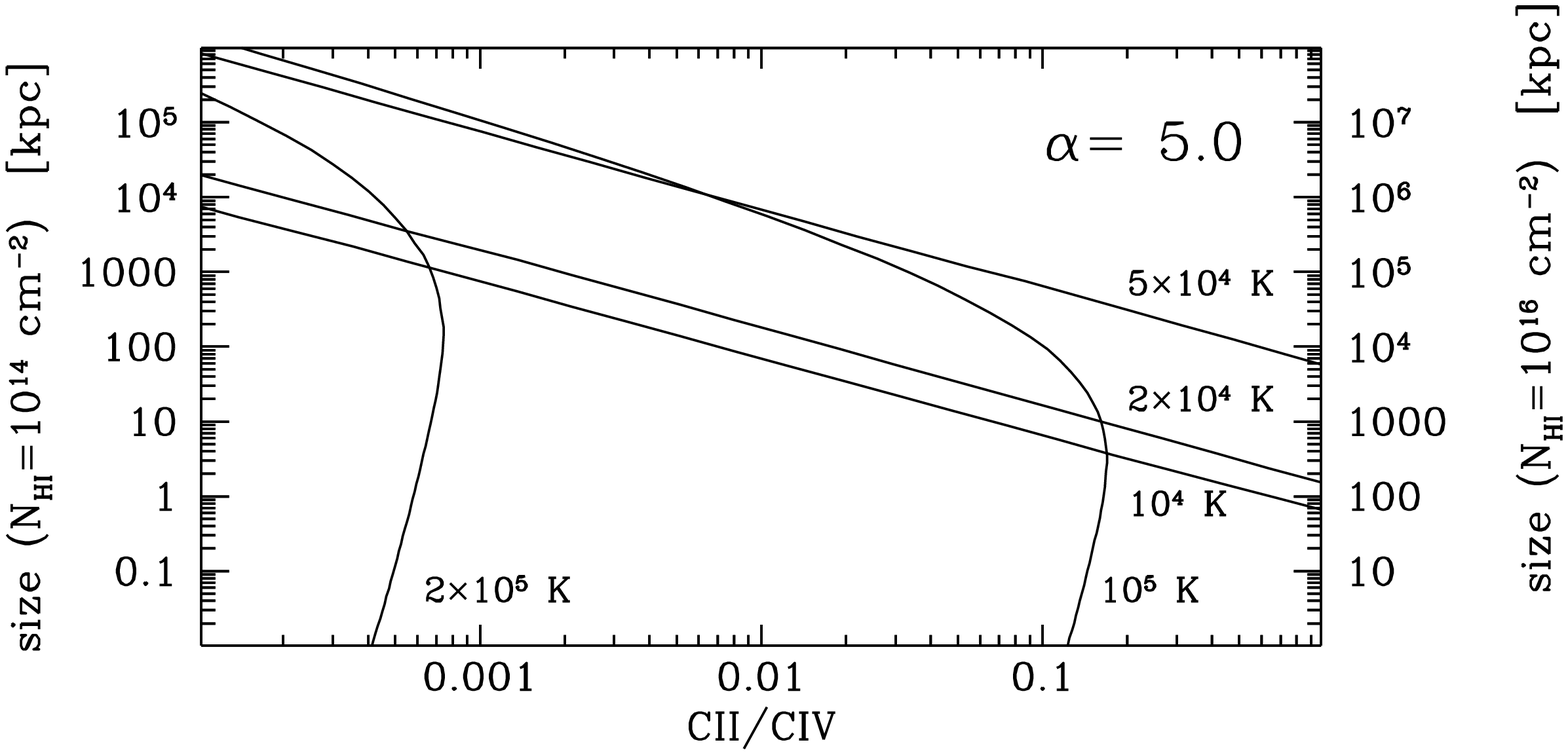,width=14.cm}

\vspace{-2.0cm}
\caption{The inferred extent parallel to  the line-of-sight 
is shown for an absorption system with neutral hydrogen 
column densities of $10^{14} \cm^{-2}$ (left axis)  and 
$10^{16} \cm^{-2}$ (right axis) as a function of the CII/CIV ratio.  
The solid curves assume the  absorber to be photo-ionized 
($I_{\nu} \propto \nu^{-\alpha}$) and to have 
a temperature and spectral index $\alpha$ as indicated on the plot. 
The dashed  curve assumes that  the temperature takes the
equilibrium value where photo-heating balances line cooling.
The upper panel is for $\alpha =1.5$, the lower panel is 
for $\alpha = 5.0$.}
\end{figure}

\end{document}